%% file: main.tex
\documentclass[12pt,a4paper]{article}

\usepackage[a4paper,margin=2.5cm]{geometry}

\usepackage{newtxtext,newtxmath} 

\usepackage{microtype}
\usepackage{setspace}
\usepackage{titlesec}
\titleformat{\section}
  {\normalfont\bfseries\fontsize{12}{14}\selectfont}{\thesection}{1em}{}
\titlespacing*{\section}{0pt}{1.2ex plus .2ex}{0.8ex}
\titleformat{\subsection}
  {\normalfont\itshape\fontsize{12}{14}\selectfont}{\thesubsection}{1em}{}
\titlespacing*{\subsection}{0pt}{1.0ex plus .2ex}{0.6ex}

\usepackage[font=bf,size=small,justification=centering]{caption}
\usepackage{graphicx}
\usepackage{booktabs}
\usepackage{multirow}
\usepackage{siunitx} 

\usepackage[style=apa,backend=biber,sortcites=true,doi=false,isbn=false,url=false,eprint=false]{biblatex}
\DeclareLanguageMapping{english}{english-apa}
\AtEveryBibitem{
  \clearfield{note}
  \clearlist{note}

    \ifentrytype{misc}{\toggletrue{bbx:url}}{}%
  
}

\usepackage{titling}
\pretitle{\begin{center}\vspace*{-1em}\LARGE\bfseries} 
\posttitle{\par\vspace{0.5em}\end{center}}
\preauthor{\begin{center}\normalsize}
\postauthor{\par\end{center}}
\predate{}\postdate{}

\usepackage{authblk}

\usepackage[hidelinks]{hyperref}
\usepackage{svg}

\title{Explainable Forecasting of Scientific Breakthroughs from Concept Network Dynamics}

\author[1,4]{Thomas Maillart}
\author[1,4]{Thibaut Chataing}
\author[5]{Ntorina Antoni}
\author[2]{David Dosu}
\author[3]{Paul Bagourd}
\author[3]{Julian Jang-Jaccard}
\author[3]{Alain Mermoud}

\affil[1]{\texttt{Thomas.Maillart@unige.ch}, \texttt{thibaut.chataing@unige.ch}\\
Geneva School of Economics and Management, University of Geneva, Geneva, Switzerland}

\affil[2]{\texttt{david.dosu@cern.ch}\\
Open Quantum Institute, CERN, Geneva, Switzerland}

\affil[3]{\texttt{paul.bagourd.ar@gmail.com}, \texttt{julian.jang-jaccard@armasuisse.ch}\\
armasuisse Science + Technology, Switzerland}

\affil[4]{Faculty of Medicine, University of Geneva, Geneva, Switzerland}

\affil[5]{\texttt{n.antoni@tue.nl}\\
TU Eindhoven, The Netherlands}

\date{} 

\begin{document}
\maketitle

\renewcommand{\thefootnote}{}%
\footnotetext{An earlier version of this work was presented at Global Tech Mining Conference 2026 (submission~\#71). This is a revised and extended preprint.}%
\renewcommand{\thefootnote}{\arabic{footnote}}%

\begin{abstract}
\input{sections/00_abstract}
\end{abstract}

\input{sections/01_introduction}
\input{sections/02_background}
\input{sections/03_theory}
\input{sections/04_method}
\input{sections/05_results}
\input{sections/06_discussion_shorter}

\input{sections/07_conclusion}

\vspace{0.2cm}

\paragraph{Data \& Source code:} All code and the manuscript sources are openly available at \url{https://github.com/wazaahhh/breakthroughs-forecasting}. The analysis is built entirely on the open OpenAlex corpus (\url{https://openalex.org}).

\paragraph{Acknowledgments:}

T.C. and T.M. acknowledge funding from armasuisse Science + Technology. D.D. acknowledges support from Open Quantum Institute. The authors acknowledge support of Julian-Jang Jaccard and Paul Bagourd, especially regarding providing and processing the use cases. 

\vspace{0.2cm}
\printbibliography



\renewcommand{\thesection}{A.\arabic{section}}
\renewcommand{\thesubsection}{A.\arabic{section}.\arabic{subsection}}
\setcounter{section}{0}
\setcounter{figure}{0}
\setcounter{table}{0}
\renewcommand{\thefigure}{A\arabic{figure}}
\renewcommand{\thetable}{A\arabic{table}}

\clearpage
\appendix
\input{sections/appendix_transfer}

\end{document}

%% file: sections/00_abstract.tex
\noindent We introduce an explainable machine-learning approach that forecasts the structural precursors of scientific breakthroughs---the emergence and intensification of links between research concepts---by modelling how OpenAlex concept networks evolve over time. Using 59 semantic and topological features, a two-stage LightGBM model jointly predicts the formation and the future weight of concept pairs, adding a regression stage that quantifies expected intensity to prior link-existence forecasts (\cite{krenn_predicting_2020,gu_forecasting_2025}). Relative to the state of the art, the approach improves accuracy \emph{and} explainability at once: comparative validation across four technology and biomedical domains yields ROC--AUC in $[0.954,\,0.967]$ at all horizons without re-tuning, exceeding the $\sim$0.90 of prior models, while every forecast rests on structural, auditable features rather than opaque embeddings. Classification performance is high (AUC $\approx$ 0.95) and regression remains stable (RMSLE $0.45 \to 0.6$ over one to five years). Feature attribution shows that structural factors---particularly Adamic--Adar similarity and degree-based Hadamard measures---consistently drive accuracy, suggesting that breakthrough-relevant recombinations emerge in tightly connected sub-networks. Two expert-anchored cases, {\it quantum annealing} and {\it AI-enabled quantum architectures}, show the model surfacing technological convergence consistent with expert expectations. We then outline a three-layer decision architecture---detection, expert translation, institutional integration---that turns these forecasts into evidence-based research strategy and policy, anchored in open data and explainable features.

%% file: sections/01_introduction.tex
\vspace{0.5cm}
\section{Introduction}
Scientific breakthroughs increasingly emerge from combining ideas across rapidly evolving knowledge networks (\cite{uzzi_atypical_2013,sinatra_quantifying_2016}). This acceleration, amplified by open science and distributed collaboration, is reshaping how science is practised and governed (\cite{fortunato_science_2018,bornmann_are_2020}), shifting discovery from isolated insight to machine-augmented, cross-disciplinary interaction (\cite{wuchty_increasing_2007,evans_metaknowledge_2011}). These dynamics challenge traditional foresight approaches such as expert panels, Delphi methods, and bibliometric trend analyses, which were originally designed for slower and more predictable innovation landscapes (\cite{martin_origins_2010,miles_dynamic_2012}). The rapid convergence of new technological innovations makes scenario-based reasoning increasingly insufficient. For example, fields such as quantum computing, AI, biotechnology, and materials science are coming together, while digital infrastructures are becoming deeply embedded in social systems (\cite{zhou_forecasting_2020,liu_mapping_2024}). As technological transitions accelerate and is increasingly tied to national competitiveness and sovereignty, governments and industries must adapt their strategies and investments in near real time (\cite{bradford_digital_2023}).

This context defines a new frontier for the science of science: explaining why some combinations of ideas lead to breakthroughs while others fade. Quantum computing exemplifies this uncertainty. Progress depends on hardware, algorithms, and co-design, particularly in the Noisy Intermediate-Scale Quantum (NISQ) era, where hybrid quantum–classical architectures and advances in error correction shape long-term viability (\cite{carleo_machine_2019,dunjko_machine_2018}). This makes quantum computing a prototypical domain in which robust and explainable foresight is essential.

Earlier studies have shown that evolving concept graphs can anticipate emerging research trajectories (\cite{krenn_predicting_2020,gu_forecasting_2025}), with link-existence ROC--AUC already exceeding 0.90. Building on this foundation, our contribution improves on the state of the art along two dimensions at once: \emph{accuracy}, raising link-existence ROC--AUC to $[0.954,\,0.967]$ across three further technology and biomedical domains without re-tuning, and \emph{explainability}, restricting the model to structural, auditable features so that every prediction and its attribution can be inspected by domain experts—rather than offering a novel forecasting principle, we deliver a more accurate, transparent, and policy-actionable specification of it. We follow the science-of-science literature in distinguishing two complementary definitions of a breakthrough: \emph{recombination-based}, where a breakthrough is the first meaningful integration of previously disconnected ideas (\cite{uzzi_atypical_2013}), and \emph{impact-based}, where a breakthrough is identified ex post by disruptive citation patterns or sustained high impact (\cite{funk_dynamic_2017,min_predicting_2021}). We forecast the \emph{structural precursors} captured by the first definition---the appearance and intensification of concept-pair links---which act as observable antecedents of the second.
Three design choices distinguish our framework from prior link-prediction work. First, we exploit the \emph{OpenAlex curated concept hierarchy}: every node is a controlled, semantically traceable concept rather than a learned embedding or raw n-gram, which makes both predictions and feature attributions auditable by domain experts. Second, we add a \emph{regression stage} on top of link-existence prediction, so the pipeline forecasts not only \emph{whether} a concept pair will form but \emph{how strongly} it will be co-cited---a quantity policy actors need to size investments. Third, we forecast on \emph{structural, explainable features only} (Adamic--Adar, degree Hadamard, weighted Hadamard, etc.), enabling per-edge feature attribution that ports directly into the decision architecture proposed in Section~\ref{sec:discussion}. Predictive performance remains high across horizons ($AUC \approx 0.95$ for existence; RMSLE $0.45 \to 0.6$ at $1 \to 5$ years for weight), and the same handful of structural features dominates attribution at every horizon. Breakthrough-relevant recombinations therefore do not appear as isolated events but as outcomes of dense, cohesive subnetworks shaped by clustering and selective bridging---a finding consistent across the existence and intensity stages.

The remainder of this paper is organized as follows. Section \ref{sec:related_work} reviews related literature. Section \ref{sec:theory} develops a network-based theory of breakthroughs. Section \ref{sec:methods} presents the data and forecasting pipeline. Section \ref{sec:results} reports predictive performance and case studies. Section \ref{sec:discussion} discusses implications for governance and introduces a three-layer decision architecture. Section \ref{sec:conclusion} concludes with perspectives for policy.

%% file: sections/02_background.tex
\vspace{0.5cm}
\section{Related Work}
\label{sec:related_work}

Mapping scientific knowledge through citation, co-authorship, and co-occurrence networks has long been central to scientometrics (\cite{perc_growth_2010,chen_citespace_2016,fortunato_science_2018}). Foundational tools such as {\it CiteSpace} enabled visualization of structural and temporal shifts in scientific domains, revealing emergent research fronts and disciplinary reconfigurations. Later studies (\cite{bornmann_convergent_2021}) extended these approaches to examine how national innovation systems and disciplinary clusters evolve through collaboration and citation patterns. Despite their descriptive richness, such methods remain largely retrospective, offering limited capacity to anticipate how new ideas or technologies will emerge. Recent advances in machine learning have shifted scientometrics from mapping to forecasting. Representation learning and graph-based models now capture latent semantic and structural relations among scientific ideas. {\it SPECTER} introduced transformer-based embeddings that reveal cross-disciplinary semantic proximity (\cite{cohan_specter_2020}). Building on this foundation, \cite{gu_forecasting_2025} as well as \cite{krenn_predicting_2020} demonstrated that evolving knowledge graphs can predict high-impact research topics before they materialize. Other work (\cite{ma_scientific_2018,meisenbacher_improved_2024,wang_exploration_2024}) linked combinatorial novelty to long-term impact, while deep learning and GNN-based approaches  expanded predictive capabilities across domains (\cite{behrouzi_predicting_2020,xu_scientific_2022}). Yet these models often prioritize accuracy over explainability.

Conceptually, the science-of-science literature offers two complementary operationalizations of a scientific breakthrough. \emph{Impact-based} definitions identify breakthroughs ex post through citation dynamics or disruption-style indicators (\cite{funk_dynamic_2017,min_predicting_2021}). \emph{Recombination-based} definitions identify breakthroughs as the integration of previously disconnected ideas (\cite{uzzi_atypical_2013,fleming_recombinant_2001}), making them observable in concept or knowledge graphs before any citation signal accrues. The two views are not in conflict: recombination events are structural antecedents, of which only a subset later acquires breakthrough-level impact. Our framework targets the antecedent layer---where actionable foresight has the most lead time---while remaining compatible with downstream impact validation through bibliographic indicators that share the OpenAlex primary key.

Parallel developments in technological forecasting mirror this trend. Graph-based and semantic methods increasingly anticipate technology convergence and diffusion (\cite{zhou_forecasting_2020}). Hybrid indicators combining publications, patents, and topic models provide early signals of emerging trajectories (\cite{liu_mapping_2024}), while large language models detect weak signals through semantic novelty and temporal embeddings (\cite{cohan_specter_2020}). Beyond predictive models, long-standing structural network metrics, such as centrality, clustering, neighbourhood connectivity, have been consistently associated with innovation potential (\cite{uzzi_atypical_2013,bornmann_are_2020,wang_exploration_2024}). Empirical evidence links breakthrough probability to balanced combinations of conventionality and novelty, cohesive local structures with bridging nodes, and persistent meso-scale configurations (\cite{sinatra_quantifying_2016,fortunato_science_2018}). These insights provide an explainable foundation for forecasting: the architecture of knowledge networks encodes early signals of conceptual recombination. Despite this progress, most state-of-the-art forecasting approaches emphasise predictive accuracy through complex and/or opaque architectures or proprietary representations, limiting insight into the mechanisms that generate predicted shifts. This opacity constrains both theory-building and policy use, as institutions have limited ability to interpret predictions or translate them into testable hypotheses and actionable portfolio choices. By aligning computational methods with transparency and reproducibility, the science of science is moving toward frameworks that not only predict innovation but also elucidate its structural logic.

%% file: sections/03_theory.tex
\vspace{0.5cm}
\section{Theoretical Approach}
\label{sec:theory}

\begin{figure}[h!]
    \centering
    \includegraphics[width=0.90\linewidth]{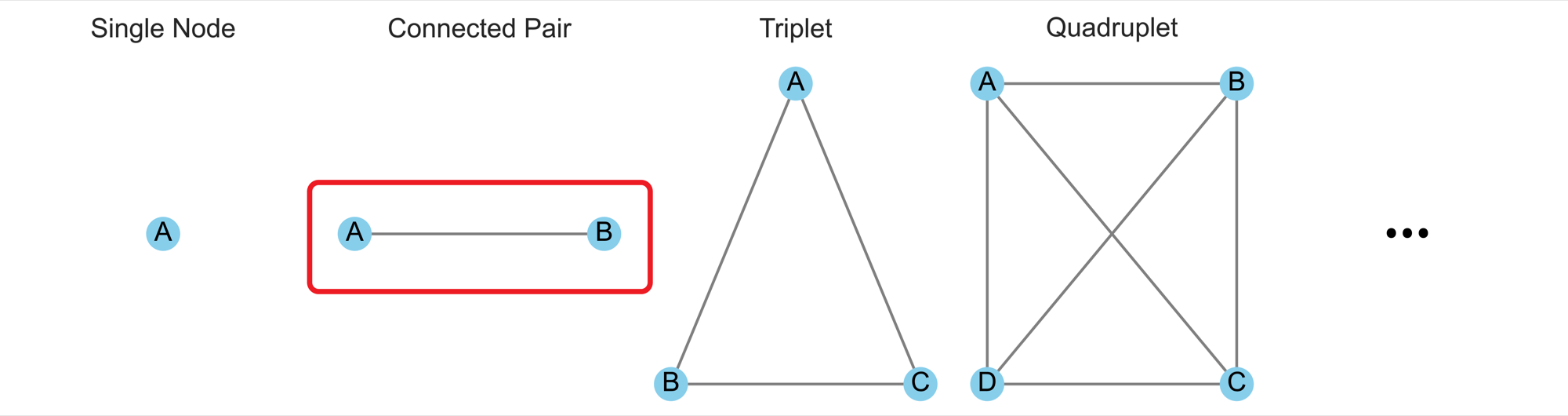}
    \caption{\footnotesize Various possibilities for structural change in the concept graph. Breakthroughs can appear as new nodes, new connected pairs, triplets, quadruplets, etc. or even more specific structures. Here, we consider the formation and the evolution of pairs to gauge for breakthroughs, as they are simple enough and appear sufficiently often in the graph to allow efficient machine learning.}
    \label{fig:breakthrough_structures}
\end{figure}

Following the recombination tradition in the science of science (\cite{uzzi_atypical_2013,fleming_recombinant_2001,perc_self-organization_2013}), we treat breakthroughs as endogenous outcomes of how the concept network rewires itself, where semantic concepts are {\it nodes} and empirical co-occurrence in the literature defines {\it weighted edges}. We are explicit about what our model forecasts and what it does not: we do \emph{not} forecast a breakthrough as defined by retrospective bibliometric impact (e.g.\ disruption indices or citation percentiles, \cite{funk_dynamic_2017,min_predicting_2021}); rather, we forecast its \emph{structural antecedent}---the appearance and intensification of specific network configurations that, in the recombination view, are necessary conditions for a breakthrough to occur. While new concept nodes mark the introduction of entirely novel concepts, new edges capture the first meaningful integration of previously independent ideas (e.g., {\it quantum computing} with {\it machine learning}). Higher-order structures such as triplets, quadruplets or more sophisticated structures may reflect clusters of coalescing concepts, signalling the consolidation of subfields or the emergence of hybrid disciplines (Figure \ref{fig:breakthrough_structures}). Recombination events can thus appear as atomic edges or as increasingly complex configurations, depending on the scale at which conceptual recombination unfolds. Hence, from a graph perspective, our theoretical approach is flexible regarding what one might consider as a breakthrough precursor. Here, we use the formation of a new link between previously weakly connected concepts as a minimal, observable unit of conceptual recombination, whose subsequent evolution determines whether it remains transient or propagates into a sustained research trajectory. The empirical link between such precursors and downstream impact (citation, disruption) is the natural next validation step and is discussed in Section~\ref{sec:discussion}.

\begin{figure}
    \centering
    \includegraphics[width=.9\linewidth]{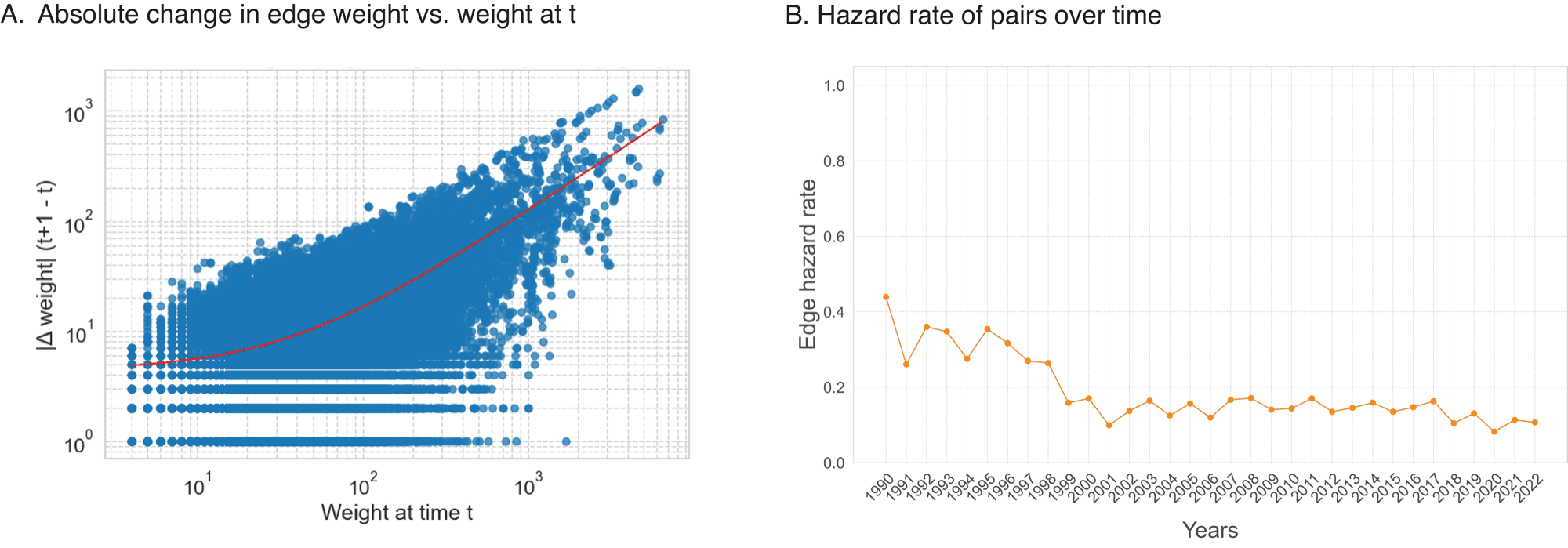}
    \caption{A. Evidence of proportional growth in edge weights. B. Evidence of significant hazard rate (disappearance) of concept pairs, with the lowest possible threshold 1 (i.e., at least one paper in one year).}
    \label{fig:prop_growth_hazard_rate}
\end{figure}

Here, two assumptions are central to identifying breakthrough candidates. First, breakthroughs rarely arise ``out of the blue'': their precursors are latent in the network structure. A key theoretical question is therefore which structural motifs predict their emergence. Second, once a structure appears, it must survive in a highly competitive Schumpeterian environment, characterized by multiplicative stochastic proportional growth (\cite{maillart_empirical_2008}) as shown in Figure \ref{fig:prop_growth_hazard_rate}A, and a high hazard rate of disappearance (Figure \ref{fig:prop_growth_hazard_rate}B), consistent with cycles of creative destruction (\cite{schumpeter_capitalism_2013}). Identifying breakthroughs thus requires understanding both their formation and their persistence. The probability that a structural change becomes a breakthrough depends jointly on local topology and global diffusion dynamics. 

To operationalize this theory, we focus on the formation and evolution of concept pairs within the quantum computer sub-graph of OpenAlex (\cite{priem_openalex_2022}). Concept pairs represent the smallest observable unit of conceptual recombination and occur frequently enough to support robust statistical analysis. Tracking how new pairs appear and how their edge weights evolve, measured by annual co-occurrence in publications, reveals the ``micro-dynamics'' of idea combination. This enables (i) early detection of emerging linkages, (ii) assessment of their likelihood of growth or decay, and (iii) a mapping of how local interactions scale to the macro-level evolution of scientific domains. Together, Figures \ref{fig:breakthrough_structures} and \ref{fig:prop_growth_hazard_rate} illustrate these principles: breakthroughs correspond to distinct structural transitions whose subsequent trajectories depend on both their embeddedness in the local network and their exposure to global diffusion pressures.

%% file: sections/04_method.tex
\vspace{0.5cm}
\section{Methods}
\label{sec:methods}
We construct our dataset from OpenAlex, an open bibliographic knowledge graph indexing scholarly works, venues, institutions, and concepts (\cite{priem_openalex_2022}). The OpenAlex \emph{curated concept hierarchy} is central to our design: every node is a controlled, semantically named concept with a stable identifier and a deterministic position in a multi-level taxonomy. This has three consequences that distinguish our setup from forecasting frameworks built on learned embeddings or raw n-gram co-occurrence. First, sub-domain extraction is exact and reproducible---we operate on the full sub-tree rooted at a chosen concept rather than on a heuristically clustered corpus. Second, every prediction is \emph{semantically traceable}: the pair $(u,v)$ corresponds to two named concepts an expert can recognize, so feature attributions translate directly into auditable claims about specific conceptual links. Third, because OpenAlex shares its primary key with downstream bibliographic resources (notably SciSciNet), the same predictions can later be cross-validated against citation- or disruption-based breakthrough indicators without identifier reconciliation. Concretely, we extract the full sub-tree of concepts descending from Quantum Computer (C58053490, level $L=3$) and retrieve all works tagged with these descendants ($L \ge 3$). Publication years and concept lists are standardized to annual resolution, and only concepts with score $>0.32$ are retained to ensure semantic relevance (following OpenAlex best practices). The resulting corpus spans 1990–2023. For each year, we build a weighted, undirected concept graph in which nodes are concepts and edges connect concept pairs co-occurring in at least $q$ publications, with $q$ set to the yearly $90^{th}$ percentile and varying from $q=2$ until 1994 to $q=7$ in 2023 ($q$ does not vary wildly because as the networks grows the number of unconnected nodes increases much faster that the number of edges. Hence, most possible co-occurrences have zero weight, also explaining why we use the $90^{th}$ percentile.). This produces a consistent temporal series of graphs capturing the evolution of conceptual recombination in quantum computing.

Across all yearly graphs, we compute an extensive set of structural indicators using {\bf NetworkX} Python Library for complex network analysis. Node metrics include {\it degree} (weighted and unweighted), {\it clustering}, {\it closeness}, {\it betweenness}, {\it eigenvector}, {\it PageRank} (weighted and unweighted), {\it average neighbor degree}, {\it number of triangles}, and {\it square clustering}. For each observed edge, we compute proximity measures such as {\it Adamic–Adar}, {\it Jaccard}, {\it Resource Allocation}, {\it Preferential Attachment}, and {\it common-neighbor indices}, with weighted variants normalized across years. We engineer pairwise features by combining node attributes using {\it Absolute Differences} (asymmetry signals) and {\it Hadamard products} (synergy signals), followed by scaling and a two-stage filtering procedure to remove low-variance and highly collinear features ($|r|\ge0.95$). Prediction targets are defined for horizons $T={1,2,3,4,5}$ years; data from 2022–2023 serve as an external validation period, and remaining observations are partitioned into $90\%$ training and $10\%$ testing sets using stratified sampling based on edge-weight distributions.

\begin{figure}
    \centering
\includegraphics[width=0.8\linewidth]{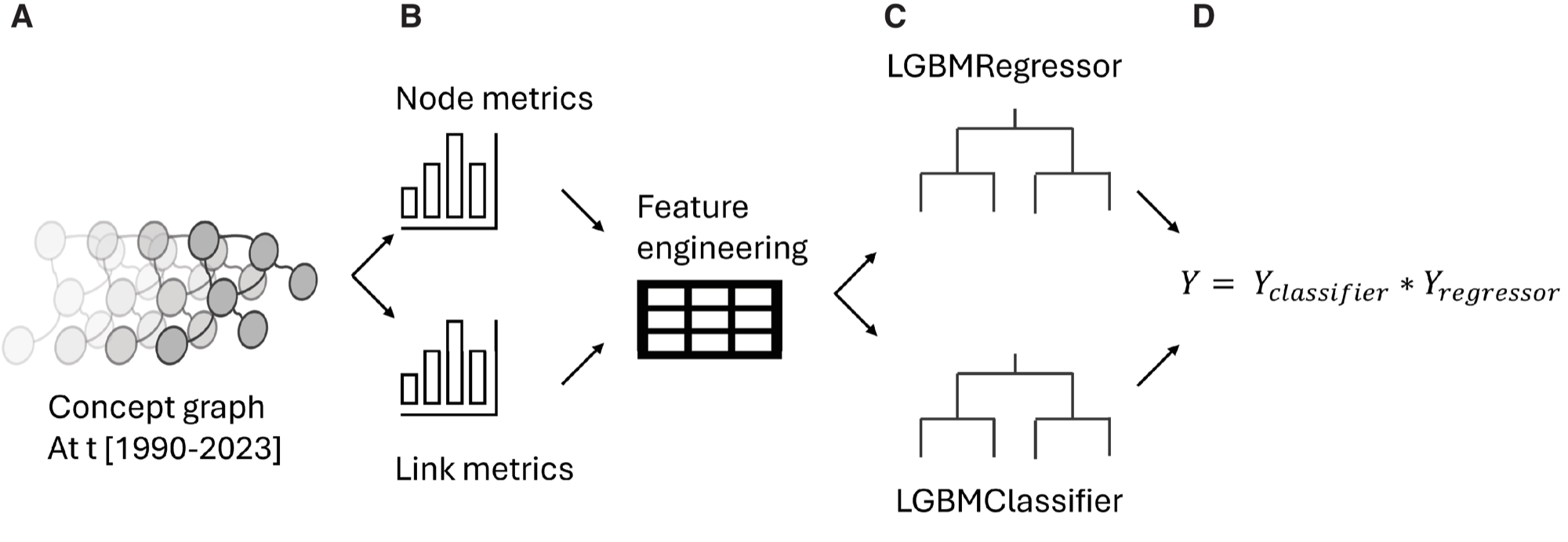}
    \caption{Machine learning pipeline: {\bf A.} The input is the evolution of the concept graph. {\bf B.} Comprehensive set of network metrics that characterize both node-level and edge-level properties. {\bf C.} Link prediction task (LGBMClassifier) and edge weight prediction task (LGRMRegressor). {\bf D.} Full prediction (link prediction + edge weight) with target function to predict edge weight within a $\pm 10\%$ tolerance range.}
    \label{fig:ML_pipeline}
\end{figure}

The forecasting pipeline (Figure \ref{fig:ML_pipeline}) follows a two-stage hurdle design. A {\it LightGBM classifier} estimates whether a concept pair will exist at horizon $T$, and a {\it LightGBM regressor} predicts its expected weight conditional on existence. The final forecast equals the product of both outputs, capturing jointly the probability and intensity of future conceptual associations. Performance is assessed using accuracy and {\it Receiver Operating Characteristic – Area Under the Curve} (ROC-AUC) for link existence, and {\it mean absolute error} (MAE), {\it Root Mean Error} (RME), and {\it Root Mean Squared Logarithmic Error} (RMSLE) to test for edge-weight prediction robustness. We further report a tolerance-based accuracy reflecting the proportion of predictions within $\pm10\%$ of the true weight. This evaluation is aligned with the stochastic multiplicative growth patterns observed in the data (Figure \ref{fig:prop_growth_hazard_rate}). Metrics are additionally stratified by logarithmic weight bins to quantify performance across a heavy-tailed distribution of conceptual linkages.

\subsection{Comparative validation protocol}
\label{sec:methods_comparative}

To test whether the structural forecasting approach depends on quantum-computing-specific graph properties, we replicated the pipeline on three additional OpenAlex concept subtrees: robotics (\emph{Robotics}, C34413123), advanced materials (nanomaterials, metamaterials, biomaterials, smart materials), and neuro implants (BCI, neuromodulation, neuroprosthetics, DBS, cochlear implant, neurostimulation). Domain definitions and seed identifiers are given in Appendix~\ref{app:domains}.

All comparative runs used an \emph{OpenAlex validation subsample}: the 30 largest non-Walden monthly partitions of the snapshot ($\approx 40\%$ of indexed works by volume, 1990--2024), preserving the relative yearly shape of the literature rather than drawing a random work sample. Absolute corpus counts in this subsample are lower than in a full snapshot by a roughly uniform factor; the primary quantum-computing results in Section~\ref{sec:results} are unchanged. Comparative metrics should be read as \emph{within-domain} robustness checks (train and test in the same subdomain), not as cross-domain transfer learning.

For the comparative study we held out all pair--year observations whose \emph{label year} (publication year plus horizon $T$) falls in 2022--2023, and trained on all earlier years with fixed LightGBM hyperparameters (no per-domain Optuna re-tuning). This protocol differs slightly from the stratified 90/10 split used for the main quantum-computing evaluation above; absolute AUC and RMSLE therefore need not match Table-level figures in Section~\ref{sec:results} digit for digit, but relative performance across domains remains informative.


%% file: sections/05_results.tex
\vspace{0.5cm}
\section{Results}
\label{sec:results}

\subsection{Pair appearance}
\begin{figure}
\centering
\includegraphics[width=0.8\linewidth]{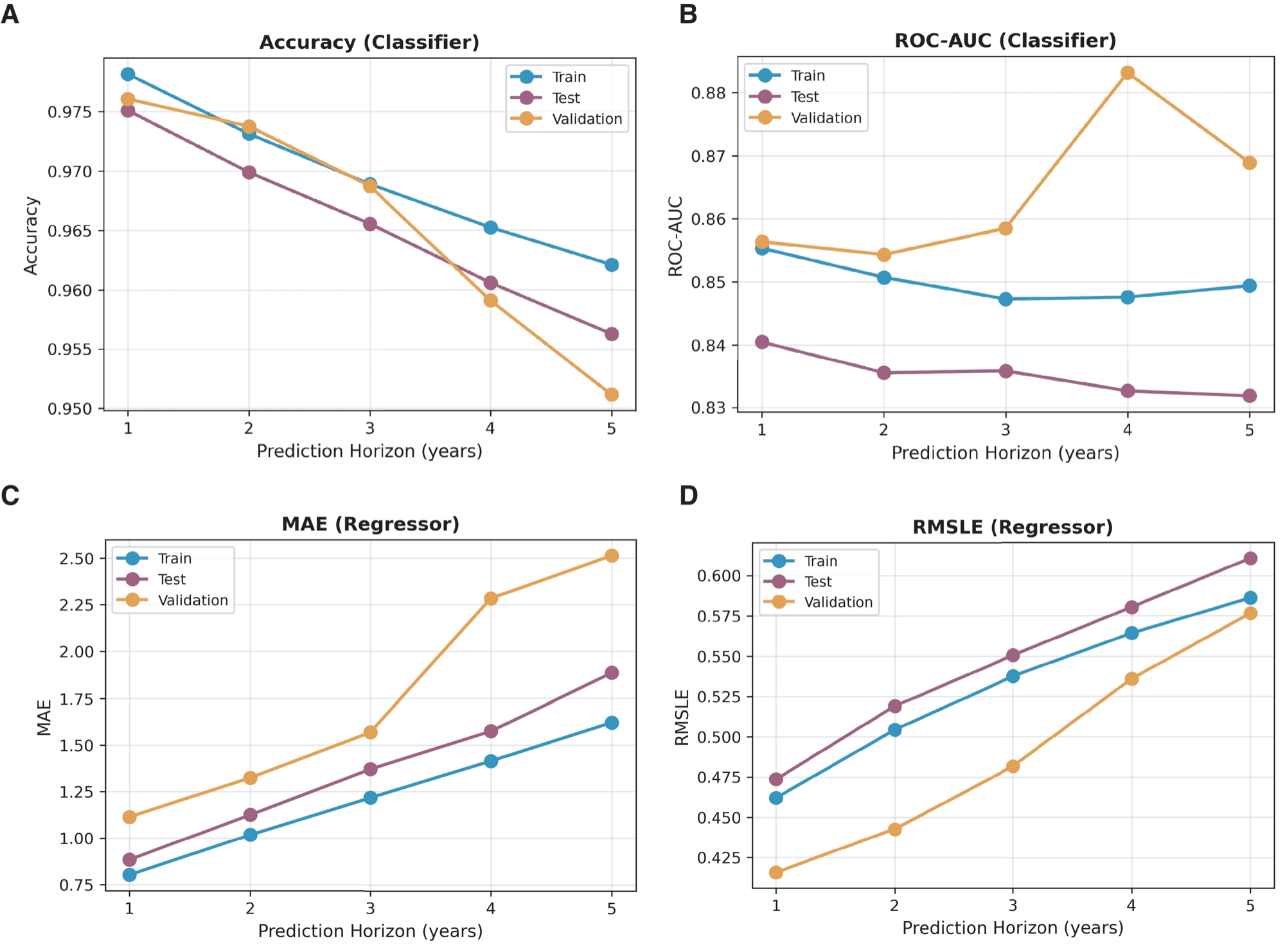}
\caption{(Upper panel) Performance of the link prediction model. {\bf A.} Accuracy decreases only marginally from $t=1$ to $t=5$. {\bf B.} ROC–AUC remains $\ge 0.95$ across all horizons. (Lower panel) Performance of the edge-strength model. {\bf C.} MAE increases slowly as $t$ grows. {\bf D.} RMSLE increases moderately (0.45 $\rightarrow$ 0.6), keeping errors within a factor of two. }
\label{fig:link_prediction}
\end{figure}

The model accurately predicts the appearance of new conceptual links in the quantum computing domain. Classification performance remains high across prediction horizons, with accuracy decreasing only slightly from $0.975$ at one year to $0.950$ at five years, while ROC–AUC remains consistently strong ($\geq 0.95$). This stability indicates that signals associated with future link formation are present in the local and mesoscopic structure of the concept network before new connections become visible in the literature. Overall, the results show that the relative likelihood of new concept pairings can be inferred from existing network configurations across multi-year horizons, supporting the use of pair-appearance prediction as a robust signal for anticipatory analysis (Figure~\ref{fig:link_prediction}).

\vspace{0.3cm}
\subsection{Regression stage}
\begin{figure}
\centering
\includegraphics[width=0.9\linewidth]{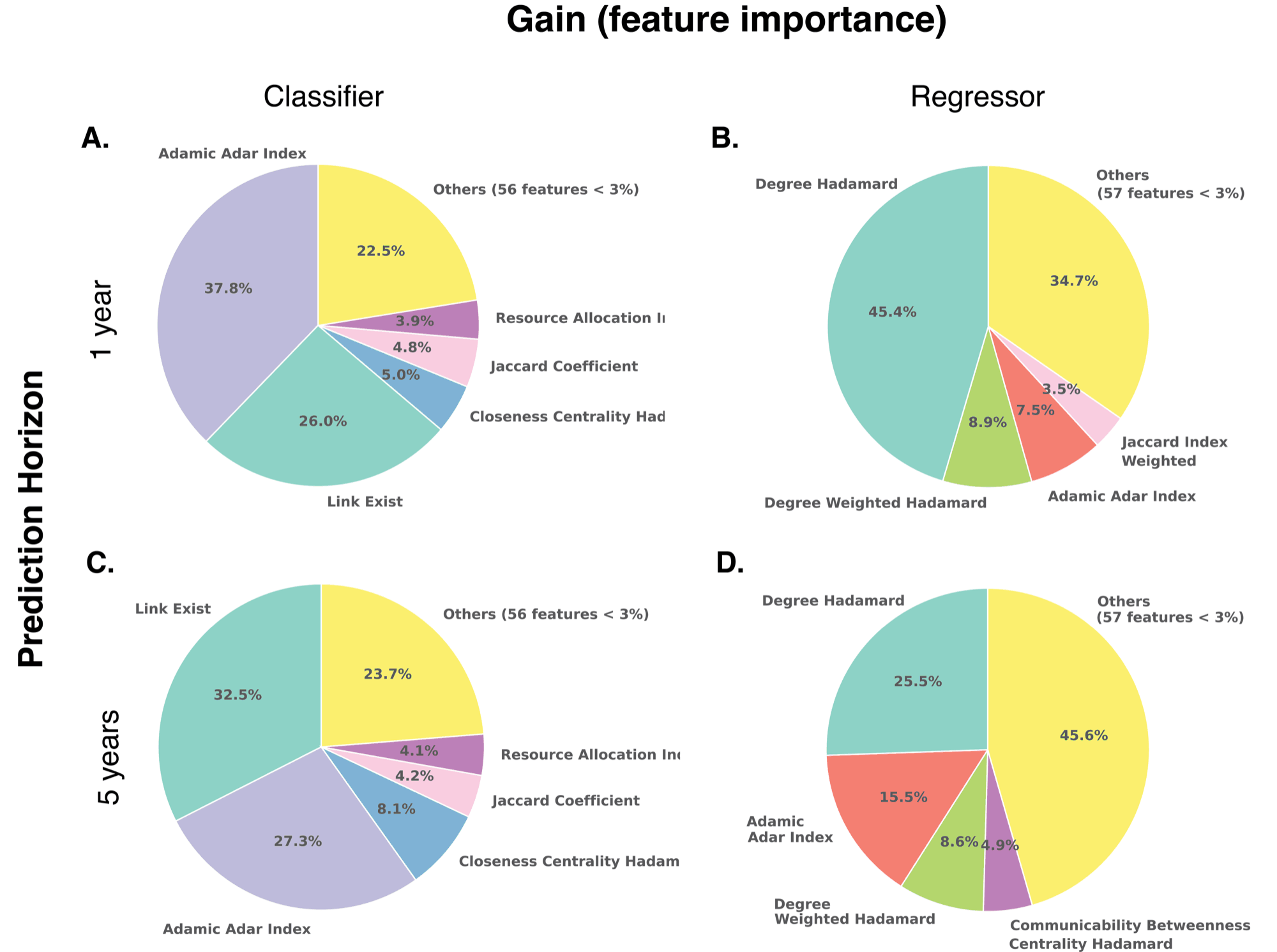}
\caption{Feature importance (gain) for prediction horizons $t=1$ and $t=5$. Adamic–Adar dominates link existence prediction; degree Hadamard dominates link-strength prediction.}
\label{fig:feature_importance}
\end{figure}

The regression stage predicting link intensity similarly shows robust performance: MAE increases slowly with the prediction horizon, and RMSLE rises from $0.45$ at one year to $0.6$ at five years, corresponding to relative deviations of $57\%$ and $82\%$ (Figure \ref{fig:link_prediction}). Predicted link weights thus remain within a factor of two of their empirical values, despite the multiplicative dynamics known to govern edge-weight growth (Figure \ref{fig:prop_growth_hazard_rate}). Feature-importance analyses reveal consistent structural mechanisms behind predictive performance (Figure \ref{fig:feature_importance}). For link prediction, the Adamic–Adar index, given by $AA = \sum_{w \in N(u) \cap N(v)} 1/\log N(w)$, where $u,v$ the two nodes for which we want to estimate link likelihood, $N(w)$ is the set of nodes adjacent to $u$ and $v$, $N(u)$ and $N(v)$ the sets of neighbours of respectively $u$ and $v$ (\cite{adamic_friends_2003}), dominates at short horizons and remains highly influential at five years, confirming the strong role of rare and specific shared neighbours. The fact that {\it link exists} becomes the strongest feature at long horizons further indicates that network evolution is highly path-dependent. For link-strength prediction, degree Hadamard, given by  $DH(u,v) = deg(u) \times deg(v)$, is the most important feature at both short and long horizons, reflecting the multiplicative capacity of well-connected concepts to amplify each other growth. Weighted Hadamard, given by $DWH(u,v) = w(u,v) \times deg(u) \times deg(v)$ further enhances this effect by incorporating edge intensity. Across both classifier and regressor, the Split metric remains diffuse (no feature $\geq 6\%$), indicating that predictions arise from a balanced set of structural cues rather than domination by a single variable.

Combining the classifier and regressor yields high end-to-end forecasting accuracy: the proportion of predictions within $\pm10\%$ of observed weights remains above $0.85$ for one-year forecasts and around $0.80$ at five years. This demonstrates that the full pipeline captures with strong reliability both the formation and the magnitude of future conceptual relations. 

\vspace{0.3cm}
\subsection{Use cases}
\label{sec:use_cases}
\begin{figure}[h!]
\centering
\includegraphics[width=1\linewidth]{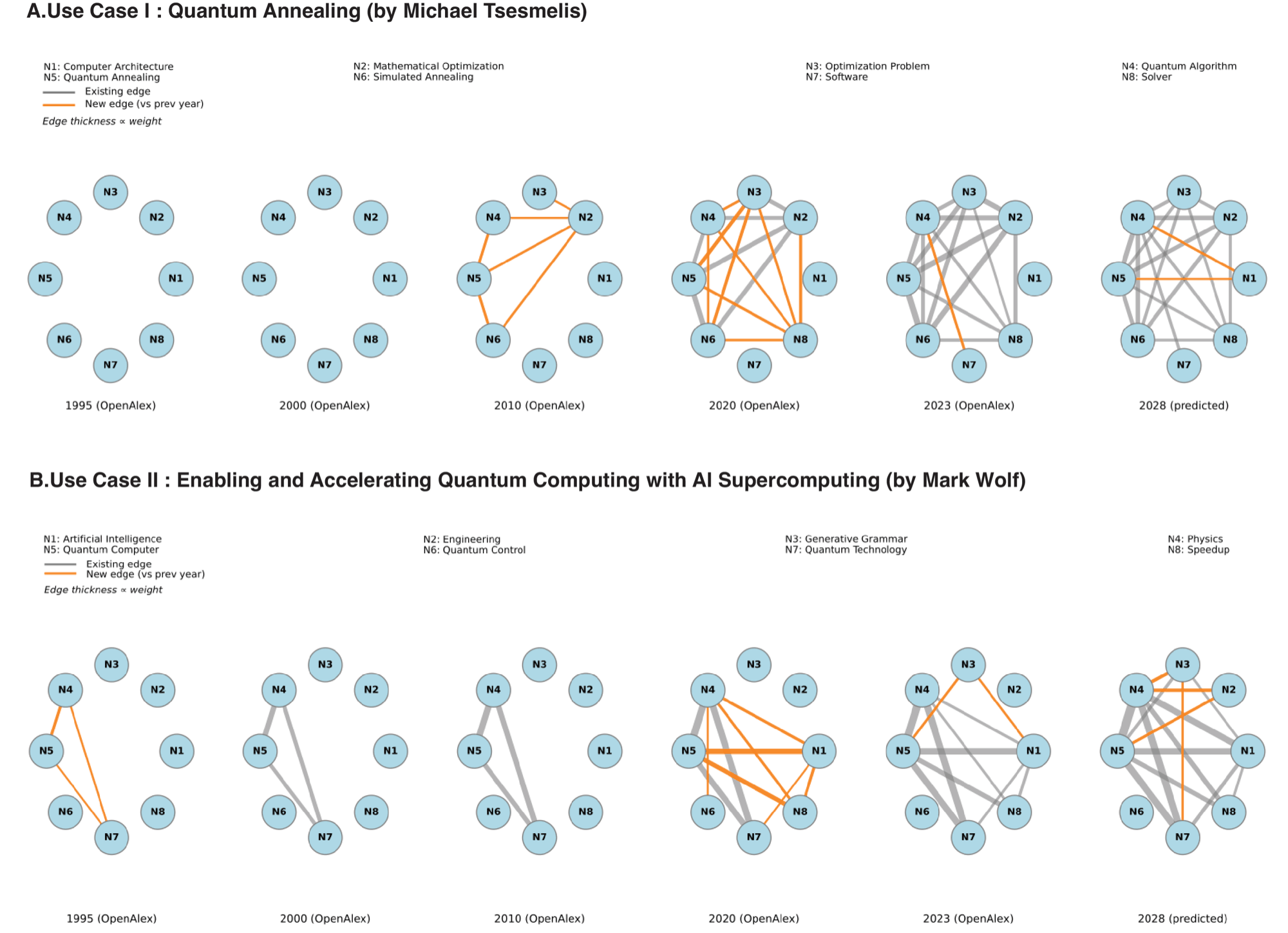}
\caption{\footnotesize Use-case validation. A: Quantum Annealing shows predicted reinforcement of core physics and optimization concepts. B: AI-accelerated quantum computing shows predicted strengthening of interdisciplinary clusters.}
\label{fig:use_cases}
\end{figure}

To validate model plausibility, we applied it to expert-defined subdomains of quantum technologies from the recently published book, {\it  Quantum Technologies - Trends and Implications for Cyber Defense} (\cite{jang-jaccard_quantum_2026}). For that, we extracted the key concepts in each of the book chapters. We then followed and predicted their evolution 5 years ahead (Figure \ref{fig:use_cases}). In Chapter 7, {\it Quantum Annealing} (Use Case I), forecasts show strengthening links between {\it Computer Architecture}, {\it Quantum Algorithm}, and {\it Quantum Annealing}, supporting expert expectations of increasing hardware–algorithm co-design. Peripheral growth in {\it Logistics Optimization} and {\it Routing Problems} indicates expansion into applied domains. In Chapter 9, {\it AI-enabled Quantum Computing} (Use Case II), the model predicts reinforced ties among {\it Engineering}, {\it Quantum Technologies}, and {\it Generative Grammar (AI \& linguistics)}, reflecting emerging convergence between machine learning, quantum control, and hybrid supercomputing. These results corroborate expert interpretations that AI is becoming a structural accelerator of quantum-technology development.


\vspace{0.3cm}
\subsection{Cross-domain robustness}
\label{sec:cross_domain}

\begin{figure}[htbp]
\centering
\includegraphics[width=0.85\linewidth]{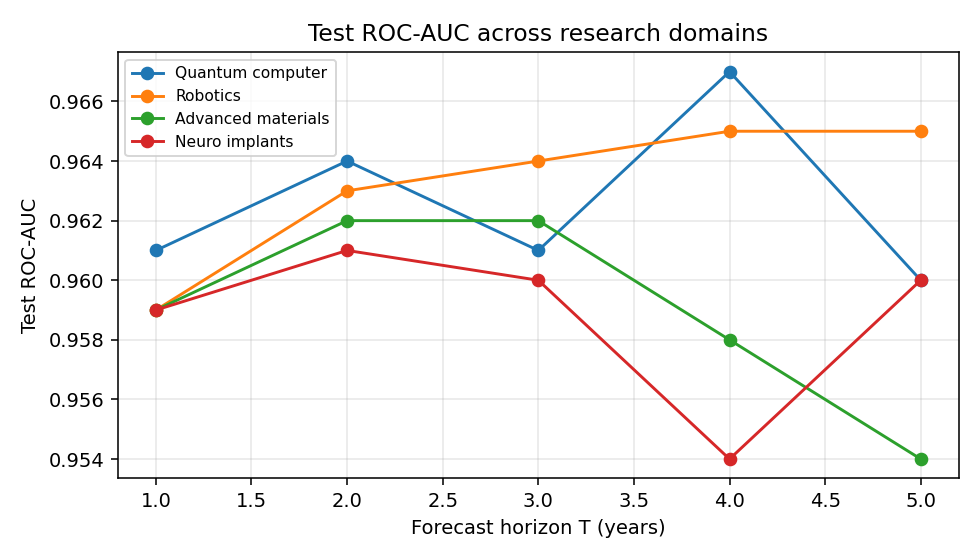}
\caption{ROC--AUC versus prediction horizon across four research domains. All domains remain in the band $[0.954,\,0.967]$ without per-domain hyperparameter tuning.}
\label{fig:comparative_auc}
\end{figure}

To test whether link-formation signals are specific to quantum-computing graph density, we replicated the full forecasting pipeline on three additional technology and biomedical subtrees (Section~\ref{sec:methods_comparative}). Table~\ref{tab:comparative_auc} summarises ROC--AUC at horizons $T=1$ and $T=5$; Figure~\ref{fig:comparative_auc} shows all five horizons.

\begin{table}[htbp]
\centering
\caption{Link-classification ROC--AUC on the OpenAlex validation subsample (test label years 2022--2023).}
\small
\begin{tabular}{lrr}
\toprule
Domain & AUC ($T=1$) & AUC ($T=5$) \\
\midrule
Quantum computer (baseline) & 0.961 & 0.960 \\
Robotics                    & 0.959 & 0.965 \\
Advanced materials          & 0.959 & 0.954 \\
Neuro implants              & 0.959 & 0.960 \\
\bottomrule
\end{tabular}
\label{tab:comparative_auc}
\end{table}

\begin{figure}[htbp]
\centering
\includegraphics[width=0.85\linewidth]{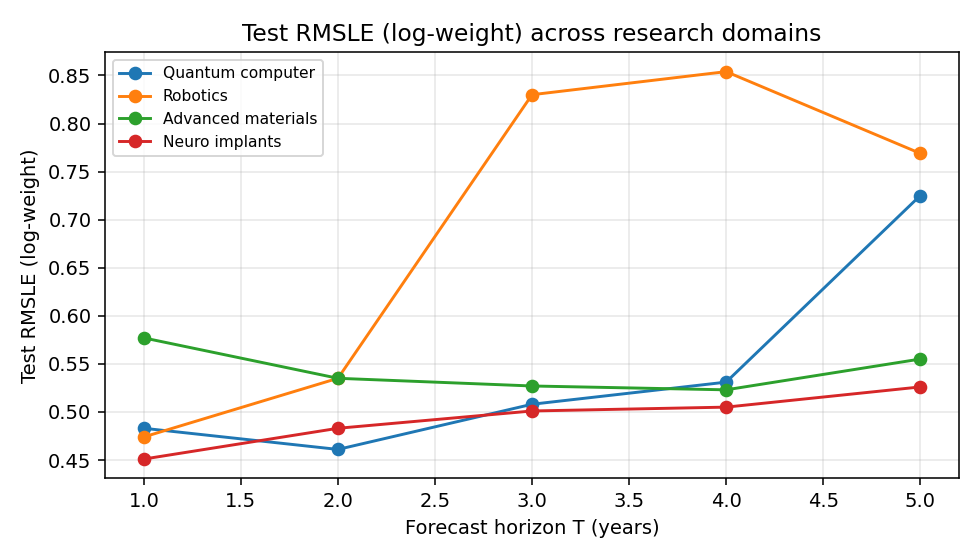}
\caption{RMSLE versus prediction horizon. Steady-growth domains (advanced materials, neuro implants) retain flat error profiles; high-volatility domains (quantum computer, robotics) degrade at longer horizons.}
\label{fig:comparative_rmsle}
\end{figure}

Classification performance is remarkably stable: every domain--horizon cell lies in $[0.954,\,0.967]$, confirming that the $\approx 50$ structural features generalise across fields with different seeding strategies and corpus growth rates. Edge-weight regression is more sensitive to domain dynamics (Figure~\ref{fig:comparative_rmsle}). Domains with moderate, steady annual corpus growth (advanced materials, neuro implants; 9--11\% per year) keep RMSLE near $0.45$--$0.55$ even at $T=5$. Domains with faster expansion (quantum computing and robotics; 15--23\% per year) show sharper RMSLE degradation at $T \ge 3$, reflecting sudden weight jumps that are harder to forecast on a log scale. The $\pm 10\%$ tolerance metric follows the same split (Appendix~\ref{app:comparative_metrics}). Together, these results indicate that the Adamic--Adar and degree-Hadamard mechanisms identified in Section~\ref{sec:results} are not artefacts of a single testbed, while weight-intensity forecasts should be interpreted jointly with a domain's growth volatility.

%% file: sections/06_discussion_shorter.tex
\vspace{0.5cm}
\section{Discussion}
\label{sec:discussion}

Our contribution addresses a persistent limitation of breakthrough forecasting, namely the opacity of many high-performing models, which constrains both theory building and policy relevance by limiting the translation of predictions into mechanisms and actionable insights (\cite{behrouzi_predicting_2020,xu_scientific_2022}). We show that concept-level network dynamics can advance predictive performance and explainability together—rather than trading one for the other, as most state-of-the-art models do—within a transparent and reproducible framework. By grounding forecasts in explicit OpenAlex concepts and explainable network features, the approach enables semantic traceability and mechanism-oriented interpretation, while remaining consistent with established accounts of recombination, consolidation, and network-driven innovation dynamics. We forecast structural \emph{precursors} of breakthroughs (link emergence and intensification) rather than retrospective impact; the natural complement is to validate predicted precursors against bibliographic impact indicators (e.g.\ disruption indices, citation percentiles) using OpenAlex-keyed resources such as SciSciNet---a step we leave for follow-up work. In bridging predictive modelling and anticipatory governance, the framework provides earlier and more actionable insight into emerging technological trajectories than standard publication- or patent-based indicators (\cite{min_predicting_2021,krenn_predicting_2020,zhou_forecasting_2020}).

\vspace{0.3cm}
\subsection{From Explainable Forecasts to Policy-Relevant Decision Architectures}
As scientific and technological change accelerates, policy choices increasingly shape long-term outcomes under deep uncertainty. In strategically sensitive and enabling domains, delayed or poorly informed intervention often proves more costly than early action based on imperfect but systematically grounded evidence, as early decisions shape trajectories and constrain future options. This reflects a core insight of anticipatory innovation governance: under high uncertainty, reactive policy responses tend to generate higher long-term costs than proactive interventions informed by structured foresight (\cite{noauthor_anticipatory_2020}). Evidence-based foresight therefore matters not because forecasts are infallible, but because they provide structured, auditable inputs that allow institutions to reason explicitly about emerging trajectories, uncertainty, and trade-offs. Translating technical forecasting capabilities into strategy and governance is thus as important as predictive performance itself. As with foresight more broadly, building future-proof, evidence-based policymaking requires decision architectures that embed forecasting into institutional routines (\cite{cuhls_foresight_2024,saritas_using_2010}). Traditional technology policy remains largely reactive, with resources and regulations adjusted only once trajectories are visible and early advantages secured (\cite{georghiou_handbook_2008,schot_three_2018}). Explainable forecasting can partially invert this sequence, but only if embedded in structured processes that transform model outputs into governance intelligence (\cite{konnola_facing_2012}). We therefore propose a three-layer decision architecture—{\bf Detection, Translation, Integration}—to convert breakthrough forecasts into anticipatory governance practice (Table~\ref{tab:three_layer_architecture}), grounded in the concrete outputs of our pipeline: predicted concept-pair links, their expected future weights, and explainable drivers of those predictions.

\textbf{Detection} establishes systematic monitoring by ranking predicted concept pairs and clusters using the probability of link existence at horizon $T$, the expected link weight conditional on existence, and explainability signals that distinguish \emph{emergence} from \emph{consolidation}. In the quantum computing use cases (Section~\ref{sec:use_cases}), increasing proximity between \textit{Computer Architecture} and \textit{Quantum Algorithms} emerged in forecast outputs before being widely recognised as a coherent research front, consistent with the co-design dynamics identified in the use-case validation (Figure~\ref{fig:use_cases}).

\textbf{Translation} combines algorithmic confidence—such as stable ROC--AUC for link existence alongside greater uncertainty for longer-horizon weight changes (Figure~\ref{fig:link_prediction})—with expert judgement to prioritise convergences. Explainability allows experts to distinguish predictions driven by nascent cross-domain recombination from those reflecting reinforcement within dense subfields, and to assess alignment with known constraints and capabilities.

\textbf{Integration} aligns validated signals with existing strategy cycles, calls, and capacity planning, enabling forecasts to inform adjustments to priorities, recruitment, and governance without creating parallel structures. Signals of consolidation can motivate coordinated, cross-disciplinary and industry-facing instruments, while early-stage emergence supports protected exploratory funding and structured monitoring across subsequent cycles. In this way, the architecture turns forecasts into repeatable portfolio routines rather than one-off analytic exercises.

\begin{table}[h]
\centering
\caption{Evidence-based decision architecture for anticipatory governance. Three layers convert breakthrough forecasts into policy action: Detection enables early awareness, Translation delivers prioritised intelligence, and Integration embeds forecasts into policy cycles through actionable readiness protocols.}
\footnotesize

\begin{tabular}{p{3cm} p{7cm} p{4.5cm}}
\toprule
\textbf{Level} & \textbf{Function} & \textbf{Outcome} \\
\midrule
\textbf{Detection} &
AI scans scientific literature, patents, and funding flows to identify emerging signals in quantum and adjacent fields. &
Early awareness of capability trajectories.\\
\textbf{Translation} &
Expert curation assesses which signals represent strategic inflection points for finance, infrastructure, and security. &
Prioritized intelligence for decision-makers. \\
\textbf{Integration} &
Framework maps findings to existing investment, policy, and planning cycles. &
Actionable readiness protocols. \\
\bottomrule
\end{tabular}

\label{tab:three_layer_architecture}
\end{table}

This architecture also creates a verifiable record of responsible governance, limiting ``we could not have known'' narratives. Integrating systematic forecasting into documented decision processes demonstrates that weak signals were monitored, uncertainty–duty-of-care trade-offs were explicitly considered, and strategic choices were grounded in structured foresight rather than ad hoc reaction. Embedding foresight into leadership practice further requires analytical literacy and anticipatory capacity: understanding technological trajectories, institutionalising foresight as policy infrastructure, and integrating early-signal monitoring into recurrent decision cycles (\cite{georghiou_handbook_2008,schot_three_2018,konnola_facing_2012}).

The European Union illustrates this logic. Quantum technologies have been positioned as a strategic priority through the 10-year Quantum Technologies Flagship launched in 2018 (\cite{european_commission_2030_2021}), commonly described as a one-billion-euro initiative, and through the integration of quantum-computing milestones into the Digital Decade agenda, including a target for a first European computer with quantum acceleration by 2025 (\cite{european_commission_quantum_2025}). At the same time, recent EU strategy highlights persistent fragmentation across Member States and funding instruments, raising risks of duplication and inefficient allocation. This creates a concrete portfolio challenge: allocating scarce resources across uncertain, fast-moving trajectories while ensuring coordination between complementary tracks, such as secure quantum communication infrastructures and computing or simulation ecosystems. In this context, explainable forecasting can act as a policy-relevant early-warning system by surfacing weak signals of promising concept convergences over multi-year horizons, enabling earlier strategic adjustment in reviews, calls, and coordination mechanisms.

\vspace{0.3cm}
\subsection{Limitations and Reproducibility}
Four limitations bound the present results. First, the framework forecasts \emph{structural precursors} of breakthroughs (link emergence and intensification) rather than retrospective bibliometric impact; whether predicted precursors disproportionately give rise to high-disruption or high-citation works is an empirical question that should be tested by joining our predictions to OpenAlex-keyed impact data (e.g.\ SciSciNet's CD-index and citation percentiles). Second, the primary quantum-computing evaluation uses the full in-paper protocol (stratified split, tuned hyperparameters); the comparative validation in Section~\ref{sec:cross_domain} uses an OpenAlex validation subsample ($\approx 40\%$ of snapshot volume), fixed hyperparameters, and a 2022--2023 label-year holdout---so absolute counts and RMSLE need not match the main results digit for digit. Third, the four-domain study tests \emph{within-domain} robustness only; it does not train on one field and predict another, and it remains complementary to external graph-based benchmarks (\cite{gu_forecasting_2025,krenn_predicting_2020}). Fourth, predictions inherit any biases in OpenAlex concept tagging---particularly at lower concept scores or for very recent works---so the score threshold ($>0.32$) and the $90^{\text{th}}$-percentile co-occurrence cut-off are conservative but not neutral choices. To support reproducibility, we use exclusively open data (OpenAlex), open-source tooling (NetworkX, LightGBM), and a fixed feature set whose construction is fully described in Section~\ref{sec:methods}.

%% file: sections/07_conclusion.tex
\vspace{0.5cm}
\section{Conclusion}
\label{sec:conclusion}
Our results indicate that the evolution of scientific knowledge networks exhibits substantial structural regularities that can be captured by explainable, feature-based machine-learning models. Focusing on concept-pair dynamics in quantum computing, we show that the structural precursors of breakthroughs follow identifiable patterns of network formation, where clustering, connectivity, and balanced novelty shape long-term trajectories. High predictive performance across time horizons ($AUC \approx 0.95$ for link existence; stable RMSLE for link weight)—improving on prior link-prediction baselines in both accuracy and explainability while relying solely on auditable structural features—highlights the value of explainable AI for scientific foresight. More broadly, the framework demonstrates that transparent model architectures and explainable features can jointly deliver prediction and explanation, enabling early detection of emergent recombination signals and the identification of structural or institutional constraints on innovation. Integrated into decision architectures such as the three-layer governance model, these methods provide a scalable basis for evidence-based foresight, linking early signals to strategic investment and institutional action. Beyond quantum computing, the approach is readily transferable to other fast-evolving domains, offering a reproducible foundation for proactive and accountable innovation governance.

%% file: sections/appendix_transfer.tex
\section{Comparative validation across four research domains}
\label{app:transfer}

\subsection{Domain definitions and OpenAlex concept seeds}
\label{app:domains}

Table~\ref{tab:app_domains} lists the four OpenAlex concept subtrees used in the comparative validation study (Section~\ref{sec:cross_domain}). Each corpus is built by recursive descent on the curated concept hierarchy from the listed seed identifiers, retaining works tagged with any descendant concept (score $>0.32$). Corpus sizes refer to the OpenAlex validation subsample described in Section~\ref{sec:methods}.

\begin{table}[htbp]
\centering
\caption{Research domains, concept seeds, and corpus characteristics on the OpenAlex validation subsample. Ann.\ growth: annual corpus growth rate, 1990--2023.}
\small
\begin{tabular}{p{2.6cm}p{4.8cm}rrr}
\toprule
Domain & OpenAlex seed concept(s) & Subtree & Works & Ann.\ growth \\
\midrule
Quantum computer (baseline) & C58053490 (\emph{Quantum computer}, L3) & 3 & 31\,935 & +22.7\,\% \\
Robotics & C34413123 (\emph{Robotics}, L3) & 6 & 51\,443 & +15.2\,\% \\
Advanced materials & C138631740 (Nanomaterials), C110367647 (Metamaterial), C2778414984 (Biomaterial), C88484716 (Smart material) & 9 & 104\,707 & +11.3\,\% \\
Neuro implants & C173201364 (BCI), C2780375056 (Neuromodulation), C197525751 (Neuroprosthetics), C2778542668 (DBS), C2778882171 (Cochlear implant), C2776443511 (Neurostimulation) & 12 & 70\,883 & +9.4\,\% \\
\bottomrule
\end{tabular}
\label{tab:app_domains}
\end{table}

\subsection{Full comparative validation metrics}
\label{app:comparative_metrics}

Table~\ref{tab:app_comparative_full} reports link-classification and edge-weight regression metrics for all four domains and horizons $T=1,\ldots,5$, evaluated on label years 2022--2023 with fixed LightGBM hyperparameters (Section~\ref{sec:methods}).

\begin{table}[htbp]
\centering
\caption{Comparative validation: full metrics by domain and horizon.}
\scriptsize
\setlength{\tabcolsep}{3pt}
\begin{tabular}{llrrrrrrr}
\toprule
Domain & $T$ & AUC & Acc. & MAE & RMSLE & $\pm10$\% & $n_{\text{train}}$ & $n_{\text{test}}$ \\
\midrule
\multirow{5}{*}{Quantum computer}
  & 1 & 0.961 & 0.885 & 12.50 & 0.483 & 0.150 & 110\,260 & 17\,348 \\
  & 2 & 0.964 & 0.895 & 11.09 & 0.461 & 0.158 & 98\,274 & 16\,074 \\
  & 3 & 0.961 & 0.892 & 14.06 & 0.508 & 0.136 & 91\,966 & 14\,204 \\
  & 4 & 0.967 & 0.900 & 14.76 & 0.531 & 0.140 & 83\,684 & 13\,316 \\
  & 5 & 0.960 & 0.871 & 23.30 & 0.725 & 0.112 & 79\,740 & 10\,356 \\
\midrule
\multirow{5}{*}{Robotics}
  & 1 & 0.959 & 0.878 & 9.76 & 0.474 & 0.140 & 204\,746 & 31\,262 \\
  & 2 & 0.963 & 0.876 & 11.95 & 0.535 & 0.130 & 186\,650 & 27\,046 \\
  & 3 & 0.964 & 0.854 & 21.19 & 0.830 & 0.078 & 176\,114 & 21\,584 \\
  & 4 & 0.965 & 0.853 & 24.30 & 0.854 & 0.086 & 164\,810 & 17\,624 \\
  & 5 & 0.965 & 0.872 & 21.11 & 0.769 & 0.086 & 153\,606 & 17\,046 \\
\midrule
\multirow{5}{*}{Advanced materials}
  & 1 & 0.959 & 0.870 & 12.88 & 0.577 & 0.131 & 379\,894 & 59\,104 \\
  & 2 & 0.962 & 0.879 & 12.31 & 0.535 & 0.139 & 340\,710 & 54\,736 \\
  & 3 & 0.962 & 0.888 & 12.66 & 0.527 & 0.139 & 312\,434 & 52\,014 \\
  & 4 & 0.958 & 0.880 & 12.42 & 0.523 & 0.136 & 281\,540 & 54\,802 \\
  & 5 & 0.954 & 0.865 & 13.33 & 0.555 & 0.130 & 256\,226 & 53\,486 \\
\midrule
\multirow{5}{*}{Neuro implants}
  & 1 & 0.959 & 0.884 & 9.25 & 0.451 & 0.151 & 217\,592 & 31\,892 \\
  & 2 & 0.961 & 0.882 & 10.74 & 0.483 & 0.150 & 202\,338 & 27\,766 \\
  & 3 & 0.960 & 0.884 & 11.13 & 0.501 & 0.139 & 185\,180 & 27\,488 \\
  & 4 & 0.954 & 0.875 & 10.77 & 0.505 & 0.149 & 171\,000 & 27\,212 \\
  & 5 & 0.960 & 0.882 & 11.76 & 0.526 & 0.142 & 160\,800 & 24\,102 \\
\bottomrule
\end{tabular}
\label{tab:app_comparative_full}
\end{table}

\paragraph{Reproducibility.}
The comparative validation uses the same feature engineering, two-stage LightGBM hurdle model, and evaluation protocol described in Section~\ref{sec:methods}.

%% file: references.bib
@misc{european_commission_2030_2021,
	title = {2030 {Digital} compass: the {European} way for the digital decade},
	url = {https://eur-lex.europa.eu/legal-content/EN/TXT/HTML/?uri=CELEX:52021DC0118},
	author = {European Commission},
	year = {2021},
}

@misc{european_commission_quantum_2025,
	title = {Quantum {Europe} {Strategy}: {Quantum} {Europe} in a {Changing} {World}},
	shorttitle = {Quantum {Europe} {Strategy}},
	url = {https://eur-lex.europa.eu/legal-content/EN/TXT/?uri=CELEX:52025DC0363},
	language = {en},
	urldate = {2025-12-12},
	author = {European Commission},
	year = {2025},
}

@techreport{noauthor_anticipatory_2020,
	type = {{OECD} {Working} {Papers} on {Public} {Governance}},
	title = {Anticipatory innovation governance: {Shaping} the future through proactive policy making},
	shorttitle = {Anticipatory innovation governance},
	url = {https://www.oecd.org/en/publications/anticipatory-innovation-governance_cce14d80-en.html},
	language = {en},
	number = {44},
	urldate = {2026-01-08},
	month = dec,
	year = {2020},
	doi = {10.1787/cce14d80-en},
	note = {Series: OECD Working Papers on Public Governance
Volume: 44},
}

@book{jang-jaccard_quantum_2026,
	address = {Cham},
	title = {Quantum {Technologies}: {Trends} and {Implications} for {Cyber} {Defense}},
	copyright = {https://creativecommons.org/licenses/by/4.0},
	isbn = {978-3-031-90726-5 978-3-031-90727-2},
	shorttitle = {Quantum {Technologies}},
	url = {https://link.springer.com/10.1007/978-3-031-90727-2},
	language = {en},
	urldate = {2025-12-11},
	publisher = {Springer Nature Switzerland},
	editor = {Jang-Jaccard, Julian and Caroff, Philippe and Blezinger, Evan and Mulder, Valentin and Mermoud, Alain and Lenders, Vincent},
	year = {2026},
	doi = {10.1007/978-3-031-90727-2},
}

@article{min_predicting_2021,
	title = {Predicting scientific breakthroughs based on knowledge structure variations},
	volume = {164},
	issn = {0040-1625},
	url = {https://www.sciencedirect.com/science/article/pii/S0040162520313287},
	doi = {10.1016/j.techfore.2020.120502},
	language = {en},
	urldate = {2022-11-11},
	journal = {Technological Forecasting and Social Change},
	author = {Min, Chao and Bu, Yi and Sun, Jianjun},
	month = mar,
	year = {2021},
	pages = {120502},
}

@article{konnola_facing_2012,
	title = {Facing the future: {Scanning}, synthesizing and sense-making in horizon scanning},
	volume = {39},
	issn = {0302-3427},
	shorttitle = {Facing the future},
	url = {https://doi.org/10.1093/scipol/scs021},
	doi = {10.1093/scipol/scs021},
	number = {2},
	urldate = {2025-11-12},
	journal = {Science and Public Policy},
	author = {Könnölä, Totti and Salo, Ahti and Cagnin, Cristiano and Carabias, Vicente and Vilkkumaa, Eeva},
	month = mar,
	year = {2012},
	pages = {222--231},
}

@book{georghiou_handbook_2008,
	title = {The handbook of technology foresight: concepts and practice},
	publisher = {Edward Elgar Publishing},
	author = {Georghiou, Luke},
	year = {2008},
}

@article{saritas_using_2010,
	title = {Using scenarios for roadmapping: {The} case of clean production},
	volume = {77},
	issn = {0040-1625},
	shorttitle = {Using scenarios for roadmapping},
	url = {https://www.sciencedirect.com/science/article/pii/S0040162510000557},
	doi = {10.1016/j.techfore.2010.03.003},
	number = {7},
	urldate = {2025-11-12},
	journal = {Technological Forecasting and Social Change},
	author = {Saritas, Ozcan and Aylen, Jonathan},
	month = sep,
	year = {2010},
	pages = {1061--1075},
}

@incollection{cuhls_foresight_2024,
	title = {Foresight: {Fifty} years to think your futures},
	booktitle = {Systems and innovation research in transition: {Research} questions and trends in historical perspective},
	publisher = {Springer Nature Switzerland Cham},
	author = {Cuhls, Kerstin and Dönitz, Ewa and Erdmann, Lorenz and Gransche, Bruno and Kimpeler, Simone and Schirrmeister, Elna and Warnke, Philine},
	year = {2024},
	pages = {73--106},
}

@article{schot_three_2018,
	title = {Three frames for innovation policy: {R}\&{D}, systems of innovation and transformative change},
	volume = {47},
	issn = {0048-7333},
	shorttitle = {Three frames for innovation policy},
	url = {https://www.sciencedirect.com/science/article/pii/S0048733318301987},
	doi = {10.1016/j.respol.2018.08.011},
	number = {9},
	urldate = {2025-11-12},
	journal = {Research Policy},
	author = {Schot, Johan and Steinmueller, W. Edward},
	month = nov,
	year = {2018},
	pages = {1554--1567},
}

@article{adamic_friends_2003,
	title = {Friends and neighbors on the {Web}},
	volume = {25},
	issn = {0378-8733},
	url = {https://www.sciencedirect.com/science/article/pii/S0378873303000091},
	doi = {10.1016/S0378-8733(03)00009-1},
	number = {3},
	urldate = {2025-11-07},
	journal = {Social Networks},
	author = {Adamic, Lada A and Adar, Eytan},
	month = jul,
	year = {2003},
	pages = {211--230},
}

@article{maillart_empirical_2008,
	title = {Empirical {Tests} of {Zipf}'s {Law} {Mechanism} in {Open} {Source} {Linux} {Distribution}},
	volume = {101},
	url = {https://link.aps.org/doi/10.1103/PhysRevLett.101.218701},
	doi = {10.1103/PhysRevLett.101.218701},
	number = {21},
	urldate = {2021-03-28},
	journal = {Physical Review Letters},
	author = {Maillart, T. and Sornette, D. and Spaeth, S. and von Krogh, G.},
	month = nov,
	year = {2008},
	note = {Publisher: American Physical Society},
	pages = {218701},
}

@book{schumpeter_capitalism_2013,
	address = {London},
	title = {Capitalism, {Socialism} and {Democracy}},
	isbn = {978-0-203-20205-0},
	publisher = {Routledge},
	author = {Schumpeter, Joseph A.},
	month = may,
	year = {2013},
	doi = {10.4324/9780203202050},
}

@article{fleming_recombinant_2001,
	title = {Recombinant {Uncertainty} in {Technological} {Search}},
	volume = {47},
	issn = {0025-1909},
	url = {https://pubsonline.informs.org/doi/abs/10.1287/mnsc.47.1.117.10671},
	doi = {10.1287/mnsc.47.1.117.10671},
	number = {1},
	urldate = {2025-11-07},
	journal = {Management Science},
	author = {Fleming, Lee},
	month = jan,
	year = {2001},
	note = {Publisher: INFORMS},
	pages = {117--132},
}

@article{wuchty_increasing_2007,
	title = {The {Increasing} {Dominance} of {Teams} in {Production} of {Knowledge}},
	volume = {316},
	issn = {1095-9203},
	url = {http://dx.doi.org/10.1126/science.1136099},
	doi = {10.1126/science.1136099},
	number = {5827},
	journal = {Science},
	author = {Wuchty, Stefan and Jones, Benjamin F. and Uzzi, Brian},
	month = may,
	year = {2007},
	pmid = {17431139},
	pages = {1036--1039},
}

@book{bradford_digital_2023,
	title = {Digital {Empires}: {The} {Global} {Battle} to {Regulate} {Technology}},
	isbn = {978-0-19-764926-8},
	shorttitle = {Digital {Empires}},
	language = {en},
	publisher = {Oxford University Press},
	author = {Bradford, Anu},
	year = {2023},
	note = {Google-Books-ID: PwTOEAAAQBAJ},
}

@article{zhou_forecasting_2020,
	title = {Forecasting emerging technologies using data augmentation and deep learning},
	volume = {123},
	issn = {1588-2861},
	url = {https://doi.org/10.1007/s11192-020-03351-6},
	doi = {10.1007/s11192-020-03351-6},
	language = {en},
	number = {1},
	urldate = {2025-11-07},
	journal = {Scientometrics},
	author = {Zhou, Yuan and Dong, Fang and Liu, Yufei and Li, Zhaofu and Du, JunFei and Zhang, Li},
	month = apr,
	year = {2020},
	pages = {1--29},
}

@article{gu_forecasting_2025,
	title = {Forecasting high-impact research topics via machine learning on evolving knowledge graphs},
	volume = {6},
	issn = {2632-2153},
	url = {https://doi.org/10.1088/2632-2153/add6ef},
	doi = {10.1088/2632-2153/add6ef},
	language = {en},
	number = {2},
	urldate = {2025-10-16},
	journal = {Machine Learning: Science and Technology},
	author = {Gu, Xuemei and Krenn, Mario},
	month = may,
	year = {2025},
	note = {Publisher: IOP Publishing},
	pages = {025041},
}

@article{liu_mapping_2024,
	title = {Mapping and comparing the technology evolution paths of scientific papers and patents: an integrated approach for forecasting technology trends},
	volume = {129},
	issn = {1588-2861},
	shorttitle = {Mapping and comparing the technology evolution paths of scientific papers and patents},
	url = {https://doi.org/10.1007/s11192-024-04961-0},
	doi = {10.1007/s11192-024-04961-0},
	language = {en},
	number = {4},
	urldate = {2025-11-07},
	journal = {Scientometrics},
	author = {Liu, Peng and Zhou, Wei and Feng, Lijie and Wang, Jinfeng and Lin, Kuo-Yi and Wu, Xuan and Zhang, Dingtang},
	month = apr,
	year = {2024},
	pages = {1975--2005},
}

@article{sinatra_quantifying_2016,
	title = {Quantifying the evolution of individual scientific impact},
	volume = {354},
	url = {https://www.science.org/doi/10.1126/science.aaf5239},
	doi = {10.1126/science.aaf5239},
	number = {6312},
	urldate = {2022-10-11},
	journal = {Science},
	author = {Sinatra, Roberta and Wang, Dashun and Deville, Pierre and Song, Chaoming and Barabási, Albert-László},
	month = nov,
	year = {2016},
	note = {Publisher: American Association for the Advancement of Science},
	pages = {aaf5239},
}

@article{uzzi_atypical_2013,
	title = {Atypical {Combinations} and {Scientific} {Impact}},
	volume = {342},
	url = {https://www.science.org/doi/full/10.1126/science.1240474},
	doi = {10.1126/science.1240474},
	number = {6157},
	urldate = {2025-11-07},
	journal = {Science},
	author = {Uzzi, Brian and Mukherjee, Satyam and Stringer, Michael and Jones, Ben},
	month = oct,
	year = {2013},
	note = {Publisher: American Association for the Advancement of Science},
	pages = {468--472},
}

@book{chen_citespace_2016,
	address = {Hauppauge},
	series = {Computer {Science}, {Technology} and {Applications}},
	title = {{CiteSpace}: a practical guide for mapping scientific literature},
	isbn = {978-1-5361-0280-2 978-1-5361-0295-6},
	shorttitle = {{CiteSpace}},
	language = {eng},
	publisher = {Nova Science Publishers, Incorporated},
	author = {Chen, Chaomei},
	year = {2016},
}

@article{miles_dynamic_2012,
	title = {Dynamic foresight evaluation},
	volume = {14},
	issn = {1463-6689},
	url = {https://doi.org/10.1108/14636681211210378},
	doi = {10.1108/14636681211210378},
	number = {1},
	urldate = {2025-11-07},
	journal = {Foresight},
	author = {Miles, Ian},
	editor = {Calof, Jonathan L. and Smith, Jack E.},
	month = feb,
	year = {2012},
	pages = {69--81},
}

@article{martin_origins_2010,
	series = {Strategic {Foresight}},
	title = {The origins of the concept of ‘foresight’ in science and technology: {An} insider's perspective},
	volume = {77},
	issn = {0040-1625},
	shorttitle = {The origins of the concept of ‘foresight’ in science and technology},
	url = {https://www.sciencedirect.com/science/article/pii/S0040162510001307},
	doi = {10.1016/j.techfore.2010.06.009},
	number = {9},
	urldate = {2025-11-07},
	journal = {Technological Forecasting and Social Change},
	author = {Martin, Ben R.},
	month = nov,
	year = {2010},
	pages = {1438--1447},
}

@article{evans_metaknowledge_2011,
	title = {Metaknowledge},
	volume = {331},
	url = {https://www.science.org/doi/full/10.1126/science.1201765},
	doi = {10.1126/science.1201765},
	number = {6018},
	urldate = {2025-11-07},
	journal = {Science},
	author = {Evans, James A. and Foster, Jacob G.},
	month = feb,
	year = {2011},
	note = {Publisher: American Association for the Advancement of Science},
	pages = {721--725},
}

@article{carleo_machine_2019,
	title = {Machine learning and the physical sciences},
	volume = {91},
	url = {https://link.aps.org/doi/10.1103/RevModPhys.91.045002},
	doi = {10.1103/RevModPhys.91.045002},
	number = {4},
	urldate = {2025-11-07},
	journal = {Reviews of Modern Physics},
	author = {Carleo, Giuseppe and Cirac, Ignacio and Cranmer, Kyle and Daudet, Laurent and Schuld, Maria and Tishby, Naftali and Vogt-Maranto, Leslie and Zdeborová, Lenka},
	month = dec,
	year = {2019},
	note = {Publisher: American Physical Society},
	pages = {045002},
}

@article{dunjko_machine_2018,
	title = {Machine learning \& artificial intelligence in the quantum domain: a review of recent progress},
	volume = {81},
	issn = {0034-4885},
	shorttitle = {Machine learning \& artificial intelligence in the quantum domain},
	url = {https://doi.org/10.1088/1361-6633/aab406},
	doi = {10.1088/1361-6633/aab406},
	language = {en},
	number = {7},
	urldate = {2025-11-07},
	journal = {Reports on Progress in Physics},
	author = {Dunjko, Vedran and Briegel, Hans J},
	month = jun,
	year = {2018},
	note = {Publisher: IOP Publishing},
	pages = {074001},
}

@article{krenn_predicting_2020,
	title = {Predicting research trends with semantic and neural networks with an application in quantum physics},
	volume = {117},
	url = {https://www.pnas.org/doi/full/10.1073/pnas.1914370116},
	doi = {10.1073/pnas.1914370116},
	number = {4},
	urldate = {2025-10-24},
	journal = {Proceedings of the National Academy of Sciences},
	author = {Krenn, Mario and Zeilinger, Anton},
	month = jan,
	year = {2020},
	note = {Publisher: Proceedings of the National Academy of Sciences},
	pages = {1910--1916},
}

@article{xu_scientific_2022,
	title = {A scientific research topic trend prediction model based on multi-{LSTM} and graph convolutional network},
	volume = {37},
	copyright = {© 2022 Wiley Periodicals LLC},
	issn = {1098-111X},
	url = {https://onlinelibrary.wiley.com/doi/abs/10.1002/int.22846},
	doi = {10.1002/int.22846},
	language = {en},
	number = {9},
	urldate = {2025-10-16},
	journal = {International Journal of Intelligent Systems},
	author = {Xu, Mingying and Du, Junping and Xue, Zhe and Guan, Zeli and Kou, Feifei and Shi, Lei},
	year = {2022},
	note = {\_eprint: https://onlinelibrary.wiley.com/doi/pdf/10.1002/int.22846},
	pages = {6331--6353},
}

@article{wang_exploration_2024,
	title = {An exploration method for technology forecasting that combines link prediction with graph embedding: {A} case study on blockchain},
	volume = {208},
	issn = {0040-1625},
	shorttitle = {An exploration method for technology forecasting that combines link prediction with graph embedding},
	url = {https://www.sciencedirect.com/science/article/pii/S0040162524005341},
	doi = {10.1016/j.techfore.2024.123736},
	urldate = {2025-10-16},
	journal = {Technological Forecasting and Social Change},
	author = {Wang, Liang and Li, Munan},
	month = nov,
	year = {2024},
	pages = {123736},
}

@article{behrouzi_predicting_2020,
	title = {Predicting scientific research trends based on link prediction in keyword networks},
	volume = {14},
	issn = {1751-1577},
	url = {https://www.sciencedirect.com/science/article/pii/S1751157720300456},
	doi = {10.1016/j.joi.2020.101079},
	number = {4},
	urldate = {2025-10-16},
	journal = {Journal of Informetrics},
	author = {Behrouzi, Saman and Shafaeipour Sarmoor, Zahra and Hajsadeghi, Khosrow and Kavousi, Kaveh},
	month = nov,
	year = {2020},
	pages = {101079},
}

@article{perc_self-organization_2013,
	title = {Self-organization of progress across the century of physics},
	volume = {3},
	copyright = {2013 The Author(s)},
	issn = {2045-2322},
	url = {https://www.nature.com/articles/srep01720},
	doi = {10.1038/srep01720},
	language = {en},
	number = {1},
	urldate = {2022-11-12},
	journal = {Scientific Reports},
	author = {Perc, Matjaž},
	month = apr,
	year = {2013},
	note = {Number: 1
Publisher: Nature Publishing Group},
	pages = {1720},
}

@article{ma_scientific_2018,
	title = {Scientific prize network predicts who pushes the boundaries of science},
	volume = {115},
	url = {https://www.pnas.org/doi/10.1073/pnas.1800485115},
	doi = {10.1073/pnas.1800485115},
	number = {50},
	urldate = {2022-10-11},
	journal = {Proceedings of the National Academy of Sciences},
	author = {Ma, Yifang and Uzzi, Brian},
	month = dec,
	year = {2018},
	note = {Publisher: Proceedings of the National Academy of Sciences},
	pages = {12608--12615},
}

@article{fortunato_science_2018,
	title = {Science of science},
	volume = {359},
	url = {https://www.science.org/doi/full/10.1126/science.aao0185},
	doi = {10.1126/science.aao0185},
	number = {6379},
	urldate = {2022-11-12},
	journal = {Science},
	author = {Fortunato, Santo and Bergstrom, Carl T. and Börner, Katy and Evans, James A. and Helbing, Dirk and Milojević, Staša and Petersen, Alexander M. and Radicchi, Filippo and Sinatra, Roberta and Uzzi, Brian and Vespignani, Alessandro and Waltman, Ludo and Wang, Dashun and Barabási, Albert-László},
	month = mar,
	year = {2018},
	note = {Publisher: American Association for the Advancement of Science},
	pages = {eaao0185},
}

@misc{priem_openalex_2022,
	title = {{OpenAlex}: {A} fully-open index of scholarly works, authors, venues, institutions, and concepts},
	shorttitle = {{OpenAlex}},
	url = {http://arxiv.org/abs/2205.01833},
	doi = {10.48550/arXiv.2205.01833},
	urldate = {2022-10-12},
	publisher = {arXiv},
	author = {Priem, Jason and Piwowar, Heather and Orr, Richard},
	month = jun,
	year = {2022},
	note = {arXiv:2205.01833 [cs]},
}

@article{perc_growth_2010,
	title = {Growth and structure of {Slovenia}’s scientific collaboration network},
	volume = {4},
	issn = {1751-1577},
	url = {https://www.sciencedirect.com/science/article/pii/S1751157710000362},
	doi = {10.1016/j.joi.2010.04.003},
	language = {en},
	number = {4},
	urldate = {2022-10-11},
	journal = {Journal of Informetrics},
	author = {Perc, Matjaž},
	month = oct,
	year = {2010},
	pages = {475--482},
}

@article{bornmann_convergent_2021,
	title = {Convergent validity of several indicators measuring disruptiveness with milestone assignments to physics papers by experts},
	volume = {15},
	issn = {1751-1577},
	url = {https://www.sciencedirect.com/science/article/pii/S1751157721000304},
	doi = {10.1016/j.joi.2021.101159},
	language = {en},
	number = {3},
	urldate = {2022-11-11},
	journal = {Journal of Informetrics},
	author = {Bornmann, Lutz and Tekles, Alexander},
	month = aug,
	year = {2021},
	pages = {101159},
}

@article{bornmann_are_2020,
	title = {Are disruption index indicators convergently valid? {The} comparison of several indicator variants with assessments by peers},
	volume = {1},
	issn = {2641-3337},
	shorttitle = {Are disruption index indicators convergently valid?},
	url = {https://doi.org/10.1162/qss_a_00068},
	doi = {10.1162/qss_a_00068},
	number = {3},
	urldate = {2022-11-11},
	journal = {Quantitative Science Studies},
	author = {Bornmann, Lutz and Devarakonda, Sitaram and Tekles, Alexander and Chacko, George},
	month = aug,
	year = {2020},
	pages = {1242--1259},
}

@article{funk_dynamic_2017,
	title = {A {Dynamic} {Network} {Measure} of {Technological} {Change}},
	volume = {63},
	issn = {0025-1909, 1526-5501},
	url = {http://pubsonline.informs.org/doi/10.1287/mnsc.2015.2366},
	doi = {10.1287/mnsc.2015.2366},
	language = {en},
	number = {3},
	urldate = {2022-11-11},
	journal = {Management Science},
	author = {Funk, Russell J. and Owen-Smith, Jason},
	month = mar,
	year = {2017},
	pages = {791--817},
}

@article{meisenbacher_improved_2024,
	title = {An {Improved} {Method} for {Class}-{Specific} {Keyword} {Extraction}},
	journal = {arXiv preprint arXiv:2407.14085},
	author = {Meisenbacher, S. and Schopf, T. and Yan, W. and Holl, P. and Matthes, F.},
	year = {2024},
}

@article{cohan_specter_2020,
	title = {Specter: {Document}-level representation learning using citation-informed transformers},
	doi = {10.48550/arXiv.2004.07180},
	journal = {arXiv preprint arXiv:2004.07180},
	author = {Cohan, Arman and Feldman, Sergey and Beltagy, Iz and Downey, Doug and Weld, Daniel S},
	year = {2020},
}
